\documentstyle[epsfig]{elsart}
\newcommand{\bm}[1]{ \mbox{\boldmath $#1$}  }

\begin{document}

\begin{frontmatter}
\title{Three-body structure of the low-lying $^{17}$Ne-states}

\author{E. Garrido} 
\address{ Instituto de Estructura de la Materia, CSIC, 
Serrano 123, E-28006 Madrid, Spain }

\author{D.V. Fedorov \and A.S.~Jensen}
\address{ Department of Physics and Astronomy,
        Aarhus University, DK-8000 Aarhus C, Denmark }

\date{\today}

\maketitle

\begin{abstract}
The Borromean nucleus $^{17}$Ne ($^{15}$O$ + p + p$) is investigated
by using the hyperspheric adiabatic expansion for a a three-body
system. The measured size of $^{15}$O and the low-lying resonances of
$^{16}$F ($^{15}$O$ + p$) are first used as constraints to determine
both central and spin-dependent two-body interactions. Then, the ground
state structure of $^{17}$Ne is found to be an almost equal mixture of 
$s^2$ and $d^2$ proton-$^{15}$O relative states, the two lowest excited 
states have about 80\% of $sd$-mixed components, and for the next two excited
three-body states the proton-$^{15}$O relative $s$-states do not
contribute. The spatial extension is as in ordinary nuclei. The widths
of the resonances are estimated by the WKB transmission through the
adiabatic potentials and found in agreement with the established
experimental limits.  We compare with experimental information and
previous works.
\end{abstract}

\end{frontmatter}

\par\leavevmode\hbox {\it PACS:\ } 21.45.+v, 27.20.+n, 21.10.Dr 

\section{Introduction}

Halo nuclei are weakly bound and spatially extended systems, and they
are expected to appear along the driplines, where nucleon single particle 
$s$ or $p$-states occur with sufficiently small separation energy 
\cite{han95,jen00,rii00,jen04}. 
Few-body techniques have proved to be very successful to describe
them \cite{zhu93}. The core degrees of freedom decouple from 
the ones of the halo nucleons, and a cluster description of the system
becomes then appropriate \cite{jen01}. 

Among the halo structures, the three-body Borromean bound systems (where
the two-body subsystems are all unbound) are specially challenging
\cite{zhu93,fed94,jen03}.  For nuclei the most likely three-body halo
candidates are Borromean systems with small binding energy. The reason
is that three particles with attractive short-range interactions are
more bound than only two of them. This is not necessarily
true when also a sufficiently strong repulsive Coulomb interaction or
Pauli repulsion is
present. Furthermore the core should contain an even number of the
nucleons surrounding it, because otherwise one valence nucleon would
be pulled into the core leaving at most one with the larger spatial
extension of a halo.

Along the neutron dripline the most studied three-body Borromean halos are
$^{6}$He and $^{11}$Li. Heavier neutron dripline systems present very
interesting challenges, but the interaction ingredients are often not
well established or the simple three-body structures are not directly
applicable. Examples could be $^{14}$Be, $^{19}$B, $^{22}$C, $^{31}$F
and $^{34}$Ne.  
Since the Coulomb interaction works against halo formation, on the
proton dripline the appearance of halos is less likely, and the
candidates necessarily require a relatively light core. In fact,
heavier halos than Ne are not expected \cite{rii92}. Still the degrees 
of freedom describing the core and the surrounding protons may decouple 
to some extent and a few-body treatment might be justified.

The interest in genuine two-proton decays have increased recently
especially due to the improved experimental techniques 
\cite{kry95,pfu02,gio02}. This fact has motivated different investigations
on light proton rich nuclei.  An especially interesting case is the
lightest Borromean dripline nucleus $^{17}$Ne ($^{15}$O+p+p), whose
excited resonance states are promising candidates to decay through
direct two-proton emission \cite{gui98,chr02}.  However, the structure
of $^{17}$Ne is very controversial. The available publications predict
very different content of the $s$ and $d$-waves
\cite{for01,tim96,nak98,mil97}.  Inconsistencies prevail even in a very 
recent and detailed three-body calculation of the structure of $^{17}$Ne
\cite{gri03}.  Nevertheless these models are used to predict new features 
of the delicate Thomas-Ehrman shifts between states in $^{17}$Ne and
$^{17}$N \cite{gri02}.

These inconsistencies may be related to the fact that $^{15}$O has a
non-vanishing spin which is difficult to handle consistently and
therefore sometimes simply ignored. Unfortunately the finite core spin
is a crucial ingredient in the structure of the excited resonance
states, at least if they have to be related to the measured resonances
of $^{16}$F nucleus. We therefore decided to investigate $^{17}$Ne
with an established method which is well tested on the neutron
dripline \cite{nie01}.  This method employs hyperspheric adiabatic
expansion which is especially suited to describe large and weakly
bound systems like halo nuclei \cite{gar02}. Furthermore, simultaneous
use of the complex scaling transformation converts the method into an
efficient tool to compute three-body resonances \cite{gar02b,fed03}.

The inconsistencies might perhaps be related to simplified treatments
of $^{17}$Ne as a three-body system of a core and two
protons. However, all the lowest excited states in the $^{15}$O-core
occur with positive parities between 5.2~MeV and 9~MeV with one
exception of 3/2$^-$ at about 6.2~MeV.  The ground, first and second
excited states of $^{17}$Ne occur with negative parities at -0.94~MeV,
0.34~MeV and 0.82~MeV from the two-proton threshold. These energies
are small compared to even the lowest-lying core-excited states, which
furthermore would have to be combined with proton valence states of
negative parity only present in the next shell. Thus two suppressing
effects must occur simultaneously to contribute to these states in
$^{17}$Ne. In fact these arguments also apply to the third and fourth
excited state in $^{17}$Ne although now the energies are only a factor
of two smaller than the core-excited states.

The 3/2$^-$ state at 6.2~MeV is then the most likely contributor, but
its energy is still relatively high. We can not exclude small
contributions form this core excitation but it seems unlikely that the
ground state is affected, since then the two protons should pay a
prize of combining into either one $d$-wave or two $p$-waves. Again
this is double suppression and rather unlikely. Still contributions to
the excited states of $^{17}$Ne are not as strongly excluded. In any
case such a core-excitation cannot be the origin of the
inconsistencies.

Insignificant contributions from core-excitations is equivalent to the
assumption of a structureless core. The probability distributions of
core and valence particles may still be overlapping provided the
corresponding degrees of freedom are dynamically decoupled. The
restriction to the smaller Hilbert space in a few-body treatment
necessitates a matching renormalization of the interaction achieved by
phenomenological constraints on the effective potential. The related
difficulty of antisymmetry between core and valence nucleons is
approximately accounted for by use of phase equivalent potentials 
\cite{gar99}.

In the present work we investigate in detail the characteristics of
the low-lying states in $^{17}$Ne in a three-body model.  We arrive at
a consistent picture describing simultaneously the two-body properties
of the internal subsystems and the global three-body properties.  The
contributions to the $^{17}$Ne states from the different partial waves
are carefully analyzed.  We do not encounter any sign of a break
between properties of ground and excited states as a signal of
substantial contribution of core-excitation from the possible 3/2$^-$
state. The convergence of the model and possible uncertainties from
the details of the two-body interactions are also investigated. In
this attempt to arrive at a consistent picture we shall confront our
calculations with results from previous work and if possible determine
the origin of discrepancies.  We already in \cite{gar04} discussed a
specific application to the Thomas-Ehrman shifts.

The paper is organized as follows: In section 2 we describe very
briefly the method used to compute three-body bound state wave functions 
and three-body resonances as well as the proton-proton and proton-$^{15}$O
interactions. With the three-body model and the input properly defined
we then in section 3 discuss the structure of the low-lying states of
$^{17}$Ne.  The accuracy of the results and the relation to previous
works are discussed in sections 4 and 5. We close the paper with a
summary and conclusions.

\section{\label{sec2} Basic ingredients}

An accurate description of a three-body system requires a suitable
method to solve the Faddeev (or Schr\"{o}dinger) equations and
the interactions between the constituent particles. We first
very briefly sketch the method and then we give a few details about
the most important of the two-body interactions

\subsection{The three-body method}

We use the hyperspheric adiabatic expansion method as described in
details for instance in \cite{fed94,nie01}. The three-body wave
function is a sum of three Faddeev components
$\psi^{(i)}(\bm{x}_i,\bm{y}_i)$ ($i$=1,2,3), each of them expressed in
one of the three possible sets of Jacobi coordinates
$\{\bm{x}_i,\bm{y}_i\}$. Each component is then expanded in terms of a
complete set of angular functions $\{\phi_n^{(i)}\}$
\begin{equation}
\psi^{(i)}={1\over\rho^{5/2}} \sum_n f_n(\rho) \phi_n^{(i)}(\rho,\Omega_i);
(\Omega_i\equiv\{\alpha_i, \Omega_{x_i}, \Omega_{y_i} \}), 
\label{eq1b}
\end{equation}
where $\rho=\sqrt{x^2+y^2}$, $\alpha_i=\arctan({x_i/y_i})$, $\Omega_{x_i}$, 
and $\Omega_{y_i}$ are the hyperspheric coordinates. Writing the
Faddeev equations in terms of these coordinates, and inserting
the expansions (\ref{eq1b}), the Faddeev equations can be
separated into angular and radial parts:
\begin{eqnarray}
\hat{\Lambda}^2 \phi_n^{(i)}+\frac{2 m \rho^2}{\hbar^2} V_{jk}(x_i)
\left( \phi_n^{(i)} + \phi_n^{(j)}  + \phi_n^{(k)}   \right)  =
\lambda_n(\rho) \phi_n^{(i)} &&
\label{eq2} \\
\left[ -\frac{d^2}{d\rho^2} +  \frac{2m}{\hbar^2} (V_{3b}(\rho) - E)
+ \frac{1}{\rho^2}
\left( \lambda_n(\rho)+\frac{15}{4} \right) \right] f_n(\rho) \nonumber &&\\
+ \sum_{n'} \left( -2 P_{n n'} \frac{d}{d\rho} - Q_{n n'} \right)f_{n'}(\rho)
= 0 & & \;\; \;\;  \label{eq3}
\end{eqnarray}
where $V_{jk}$ is the two-body interaction between particles $j$ and
$k$, $\hat{\Lambda}^2$ is an angular operator \cite{nie01} and $m$ is
the normalization mass. In Eq.(\ref{eq3}) $E$ is the three-body
energy, $V_{3b}$ is a three-body potential used for fine-tuning and
the functions $P_{n n'}$ and $Q_{n n'}$ are given for instance in
\cite{nie01}.

It is important to note that the set of angular functions used
in the expansion (\ref{eq1b}) are precisely the eigenfunctions of 
the angular part of the Faddeev equations. Their corresponding 
eigenvalues are denoted by $\lambda_n(\rho)$, and enter in the 
radial equations (\ref{eq3}) as an essential ingredient in
the diagonal part of the effective radial potentials:
\begin{equation}
V_{\mbox{eff}}(\rho)=\frac{\hbar^2}{2m}
    \frac{\lambda_n(\rho)+15/4}{\rho^2}+V_{3b}(\rho)
\label{eq4}
\end{equation}

The complex scaling method \cite{agu71} amounts to rotation of
the hyperradius ($\rho\rightarrow \rho e^{i\theta}$), while the
hyperangles remain unchanged \cite{fed03}.  As soon as the rotation
angle $\theta$ is larger than the argument of any given resonance,
then this resonance together with the bound states is obtained as
an exponentially decreasing solution to the coupled set of radial
equations (\ref{eq3}). The problem of matching with the non trivial
Coulomb asymptotics is thus avoided, since the resonance corresponds 
to a solution with ``bound state" asymptotics. It was shown in 
\cite{fed03} that the complex scaling method allows reliable calculations
of resonance energies, also for three-body systems with the long-range
Coulomb interaction.

\subsection{Two-body potentials}

The two-body potentials must describe the properties of the two-body
subsystems. However, unreflected reproduction of the low-lying spectra
of the two-body systems is not sufficient. The subsequent use in the
context of a specific Hilbert space could be crucial. This becomes
particularly evident when a number of degrees of freedom are frozen in
the three-body computation while simultaneously also necessary to
conserve symmetries of the effective Hamiltonian.  An appropriate
example is the spin-dependence of the assumed form of the effective
two-body potentials.  A bad choice can lead to catastrophic results
for the three-body system \cite{gar03}.

For the proton-proton interaction we use the parametrization given in
\cite{gar04,gar97}. Other choices have been tried as well but the 
three-body results do not change significantly.

For the proton-$^{15}$O interaction spin-dependent operators are
indispensable to reproduce the properties of the low-lying $^{16}$F
resonances. We parametrize by use of operators that conserve the
proton total angular momentum $\bm{j}_p$=$\bm{\ell}$+$\bm{s}_p$ and
the total two-body angular momentum $\bm{j}$=$\bm{j}_p$+$\bm{s}_c$,
where $\bm{\ell}$ is the relative orbital angular momentum, and
$\bm{s}_c$ and $\bm{s}_p$ are the core ($^{15}$O) and proton spins,
respectively.  These operators permit a clear energy separation of the
usual mean-field spin-orbit partners $\ell_{\ell+1/2}$ and
$\ell_{\ell-1/2}$. It is then possible to use a proton-core
interaction such that the low-lying states have well defined
$\ell_{j_p}$ quantum numbers, like the $d_{5/2}$ states in $^{16}$F.
This is especially important when one of these states is forbidden by
the Pauli principle, like for instance the $p_{3/2}$ waves in
$^{10}$Li \cite{gar03}.

In \cite{gar04} we introduced an $\ell$-dependent proton-core 
interaction of the form:
\begin{eqnarray}
\lefteqn{
V^{(\ell)}_{p-core}(r) = } \label{eq1} \\ &&
S_c^{(\ell)} f^{(\ell)}_c(r) +S_{ss}^{(\ell)} 
f^{(\ell)}_{ss}(r) 
\bm{s}_c \cdot \bm{j}_p -  S_{so}^{(\ell)} 
\frac{1}{r}\frac{d}{dr}
f^{(\ell)}_{so}(r) \bm{\ell} \cdot \bm{s}_{p}
 + \frac{Z_c e^2}{r} {\rm Erf}(r/b_c) \; ,
\nonumber
\end{eqnarray}
where the shapes of the central ($f^{(\ell)}_c$), spin-spin
($f^{(\ell)}_{ss}$) and spin-orbit ($f^{(\ell)}_{so}$) radial
potentials are chosen to be Woods-Saxon functions,
$1/(1+\exp{((r-b_\ell)/a)})$, with the same diffuseness $a$ in all
terms.  The proton number of the core is $Z_c$.  The error function
${\rm Erf}$ describes the proton-core Coulomb interaction of a
gaussian core-charge distribution. The range parameter
$b_c$=2.16 fm reproduces the experimental rms charge radius in
$^{15}$O of 2.65 fm.

In \cite{gar04} the parameters of the Woods-Saxon radial potentials
were adjusted to reproduce the experimental energies of the four
lowest states in $^{16}$F, and simultaneously, after switching off the
Coulomb interaction, the energies of the corresponding states in the
mirror nucleus $^{16}$N. The resulting parameters are given in
table~\ref{tab1}.

\begin{table}
\caption{\label{tab1} Radius ($b_\ell$) and strengths of the central 
($S_s^{(\ell)}$), spin-spin ($S_{ss}^{(\ell)}$), and spin-orbit
($S_{so}^{(\ell)}$) potentials in Eqs.(\ref{eq1}) and (\ref{eq5}), that
correspond to the Woods-Saxon (WS) and Gaussian (G) proton-core
interactions, respectively. In (\ref{eq1}) the diffuseness $a$
is 0.65 fm in all the cases.  }
\begin{center}
\begin{tabular}{|c|cc|cc|cc|cc|}
\hline 
$\ell$ & \multicolumn{2}{c|}{$b_\ell$ (fm)} & \multicolumn{2}{c|}{$S_c^{(\ell)}$ (MeV)}  & 
          \multicolumn{2}{c|}{$S_{ss}^{(\ell)}$ (MeV)}  & 
          \multicolumn{2}{c|}{$S_{so}^{(\ell)}$ (MeV$\cdot$fm$^2$, MeV)}  \\
\hline
  &  WS &  G  & WS &  G  & WS &  G  & WS &  G \\
\hline
 $0$ & 3.00  & 2.60 &$-53.91$ & $-76.56$ &0.92 &1.34& -- & --\\
 $1$ & 2.70  & 2.80 &$-19.99$ & $-21.42$ &0.69 &1.04& $-25.0$ & $-8.2$\\
 $2$ & 2.85  & 2.80 &$-58.45$ & $-75.89$ &0.24 &0.32& $-25.0$ & $-8.2$ \\
\hline
\end{tabular}
\end{center}
\end{table}

In order to investigate the dependence of the results on the 
two-body interactions we have constructed an additional proton-core 
potential with gaussian form factors.
Assuming that the (possibly $\ell$-dependent) range of the gaussians is
the same for central, spin-spin, and spin-orbit terms we can write the
two-body interaction as:
\begin{equation}
V^{(\ell)}_{p-core}(r) = e^{-(r/b_\ell)^2} \left( S_c^{(\ell)}+S_{ss}^{(\ell)} 
\bm{s}_c \cdot \bm{j}_p +  S_{so}^{(\ell)} \bm{\ell} \cdot \bm{s}_{p} \right).
\label{eq5}
\end{equation}
The $s$ and $d$-wave parameters are adjusted to reproduce the
experimental energies of the 0$^-$, 1$^-$, 2$^-$, and 3$^-$ states in
$^{16}$N and $^{16}$F, see table~\ref{tab1}. The accuracy is rather
good and very similar to the Woods-Saxon case \cite{gar03} as seen in
table~\ref{tab2} where both computed and measured energies are given.

The experimental energies of the second 1$^-$/2$^-$ doublet is used to
adjust the value of the $d$-wave spin-orbit parameter.  These two
states must come from the coupling of a $d_{3/2}$-wave and the 1/2$^-$
state of the core. In the same way, the measured 1$^+$/2$^+$ doublet,
coming from the coupling of a $p_{3/2}$-wave and the 1/2$^-$ state of
the core, is used to determine the $p$-wave interaction (also given in
table~\ref{tab1}).  As seen in the lower part of table~\ref{tab2}, the
computed resonance energies agree reasonably well with measurements,
but the widths are very much larger than the experimental values. The
reason is simply that the computed resonances appear above the Coulomb
and centrifugal barriers. This in turn is the result of the gaussian
form of the attractive central potential combined with the necessary
range of about $3$~fm. Then the attraction at the surface pushes the
barriers outwards and simultaneously reduces their heights.

It is possible to construct other two-body nucleon-core potentials
reproducing as well the widths of the high-lying 1$^-$/2$^-$ and
1$^+$/2$^+$ doublets. This would require a more complicated form than
Eq.(\ref{eq5}) and more than the four parameters used for each partial
wave. With the present assumptions we reproduce 6 experimental values
for $s$-waves and 10 for $d$-waves. We prefer the simplicity of
Eq.(\ref{eq5}), because components related to the relatively
high-lying $d_{3/2}$ state as well as the $p$-waves in general, as we
shall see later, never contribute by more than 4\% of the total norm
of the wave function. Another potential would essentially not change
the results as long as symmetry properties and resonance energies are
maintained, see the discussion of accuracy in section 4.

\begin{table}
\caption{\label{tab2} The two-body $^{16}$N and $^{16}$F spectra obtained 
with the gaussian potential specified in table \ref{tab1}. For unbound
states we give the corresponding energies and widths $(E_R,\Gamma)$.
The experimental error bars are not specified when they are smaller
than the last digit. For unbound states the energies are decay
energies above threshold.}
\begin{center}
\begin{scriptsize}
\begin{tabular}{|c|cc|cc|}
\hline
$J^{\pi}$ & $^{16}$N  & Exp.\cite{ajz86}  & $^{16}$F   & Exp.\cite{ajz86}  \\
\hline
 0$^-$ & $-2.37$  & $-2.371$ & (0.53,$2\cdot 10^{-3}$) & (0.535, $0.040\pm0.020$) \\
 1$^-$ & $-2.09$  & $-2.094$ & (0.70,$\sim 10^{-5}$) & ($0.728\pm0.006$, $<0.040$) \\
 2$^-$ & $-2.49$  & $-2.491$ & (0.95,0.01) & ($0.959\pm0.005$, $0.040\pm0.030$) \\
 3$^-$ & $-2.19$  & $-2.193$ & (1.21,0.01) & ($1.256\pm0.004$, $<0.015$) \\
\hline
 1$^+$ & (0.86,1.22)  & (0.86,0.02) & (3.71,3.80) & (4.29$\pm$0.01, $<0.040$) \\
 2$^+$ & (1.03,1.86)  & (1.03,$<$0.01) & (3.89,4.31) & (4.41$\pm$0.01, $<0.020$) \\
 1$^-$ & (2.33,0.96)  & (2.27$\pm$0.05,0.25$\pm$0.05) & (5.28,2.27) & (5.81$\pm$0.01,---) \\
 2$^-$ & (2.44,1.07)  & (2.56,0.02$\pm$0.01) & (5.38,2.41) & --- \\
\hline
\end{tabular}
\end{scriptsize}
\end{center}
\end{table}

The Woods-Saxon and Gaussian $s$-wave potentials in table~\ref{tab1}
have a deeply bound $^{16}$F-state at $-26.2$ MeV and $-30.5$ MeV,
respectively.  These states correspond to the $s_{1/2}$ proton states
occupied in the $^{15}$O core. They are then forbidden by the Pauli
principle, and should be excluded from the calculation.  This is
implemented as in refs.\cite{gar97,gar99} by use of the phase
equivalent potential which has exactly the same phase shifts as the
initial two-body interaction for all energies, but the Pauli forbidden
bound state is removed from the two-body spectrum.  We then use the
phase equivalent potential of the central part of the $s$-wave
potentials in table \ref{tab1}.  Thus the $s$-states actively entering
the three-body calculations are the second states of the initial
potentials.

The lowest $p$ shell is also fully occupied in $^{15}$O. We
should then apply the same treatment as for $s$-waves to the $p$-wave
proton-core interaction, using a potential with deeply bound states
that are afterwards removed by the corresponding phase equivalent
potentials.  However, since the $p$ waves basically have no effects in
the three-body calculation we use for simplicity the shallow
$\ell$=1 potential without bound states as given in table~\ref{tab1}.
This saves a substantial amount of computing time without loss of
accuracy in the calculations.

Partial waves with $\ell > 2$ are not crucial for the low-lying
states.  They should contribute on roughly the same level as the
positive parity core-excitations corresponding to energy increases
from the $p$ to the $sd$-shell, and perhaps less than the negative
parity excitation of $^{15}$O increasing the energy only by the
$p$-shell spin-orbit splitting.  Quantitative estimates can be
obtained with potentials systematically extrapolated from the $\ell
=0,1,2$ parameters in table \ref{tab1}. This can then be inferred to
indicate that core-excitations are insignificant. In any case the use
of effective potentials reproducing the pertinent measured quantities
has to be consistent with the choice of model space. Then a
phenomenologically adjusted two-body model without core-excitations
for $^{16}$F is expected to account correctly for most of the energy
in the $^{17}$Ne spectrum.

\section{Structure of the $^{17}$Ne states}

In this section we discuss the results obtained for the ground and
excited states of $^{17}$Ne. We start with the results obtained with
the Woods-Saxon (WS) proton-core potential in Eq.(\ref{eq1}).

Together with the two-body potentials we must specify the components
included in the calculations. We give them in subsection 3.1. The
output are then the effective potentials entering in the radial part
of the Faddeev equations, and the energy spectrum obtained when
solving the radial part. The potentials are briefly described in
subsection 3.2, and in 3.3 the properties of the computed $^{17}$Ne
states are analyzed in detail. We close the section by discussing the
spatial distribution of the states and provide a novel WKB estimate of
the widths of the resonances in subsections 3.4 and 3.5, respectively.

\subsection{Components}

We first compute the angular eigenvectors and eigenvalues from
Eq.(\ref{eq2}). This is done by expanding the eigenvectors in the
basis $\{ {\cal Y}_{\ell_x \ell_y, L}^{K}(\alpha_i, \Omega_{x_i},
\Omega_{y_i}) \otimes \chi_{s_x s_y, S} \}$, 
where ${\cal Y}_{\ell_x \ell_y, L}^{K}$ are the hyperspheric harmonics
and $\chi$ is the spin function \cite{nie01}. The quantum number
$\ell_x$ is the relative orbital angular momentum between particles
$j$ and $k$, and $s_x$ is the coupling of their two spins. $\ell_y$ is
the relative orbital angular momentum of particle $i$ and the center
of mass of the $jk$ two-body system, and $s_y$ is the spin of particle
$i$. The coupling of $\ell_x$ and $\ell_y$, and of $s_x$ and $s_y$,
are $L$ and $S$, respectively, which in turn couple to the total
angular momentum $J$ of the system. Finally,
$K$=$2n$+$\ell_x$+$\ell_y$, where $n$ is the usual nodal quantum
number.

\begin{table}
\caption{\label{tab3} Components included for the $1/2^{-}$
state of $^{17}$Ne. The upper part refers to the Jacobi
set where $\bm{x}$ connects the two protons. In the lower part
$\bm{x}$ connects the core and one of the protons. The sixth column
indicates the maximum value of the hypermomentum $K$. The last column
gives the contribution of the component to the total norm of the
wave function. Only those components contributing more than
1\% are given. In each Jacobi set these probabilities should then add
up to 100\%. }
\begin{scriptsize}
\begin{center}
\begin{tabular}{|cccccc|c|}
\hline
 $\ell_x$ & $\ell_y$  & $L$ & $s_x$ & $S$ & $K_{max}$ & $W(^{17}\mbox{Ne})$  \\
\hline
 0 & 0  & 0  & 0  & 1/2  & 100  & 89.0  \\
 1 & 1  & 1  & 1  & 1/2  & 52   & 2.4  \\
 1 & 1  & 1  & 1  & 3/2  & 62   & 4.7  \\
 2 & 2  & 0  & 0  & 1/2  & 54   & 3.8  \\
\hline
 0 & 0  & 0  & 0  & 1/2  & 60  &  11.4  \\
 0 & 0  & 0  & 1  & 1/2  & 80  &  34.0  \\
 2 & 2  & 0  & 0  & 1/2  & 60  &  10.4  \\
 2 & 2  & 0  & 1  & 1/2  & 84  &  31.2  \\
 2 & 2  & 1  & 0  & 1/2  & 34  &  1.9  \\
 2 & 2  & 1  & 1  & 3/2  & 44  &  4.9  \\
 1 & 1  & 0  & 1  & 1/2  & 42  &  4.1  \\
 1 & 1  & 0  & 0  & 1/2  & 42  &  1.4  \\
\hline
\end{tabular}
\end{center}
\end{scriptsize}
\end{table}

\begin{table}
\begin{minipage}[t]{7cm}
\caption{\label{tab4} As table \ref{tab3} for the $3/2^{-}$ state
of $^{17}$Ne.}
\begin{scriptsize}
\begin{center}
\begin{tabular}{|cccccc|c|}
\hline
 $\ell_x$ & $\ell_y$  & $L$ & $s_x$ & $S$ & $K_{max}$ & $W(^{17}\mbox{Ne})$  \\
\hline
 1 & 1  & 1  & 1  & 1/2  & 62   & 5.7  \\
 1 & 1  & 1  & 1  & 3/2  & 42   & 1.2  \\
 1 & 1  & 2  & 1  & 1/2  & 42   & 5.0  \\
 1 & 1  & 2  & 1  & 3/2  & 42   & 5.0  \\
 2 & 2  & 2  & 0  & 1/2  & 64   & 9.9  \\
 0 & 2  & 2  & 0  & 1/2  & 100   & 47.5  \\
 2 & 0  & 2  & 0  & 1/2  & 82   & 25.7  \\
\hline
 2 & 2  & 1  & 0  & 1/2  & 54  & 3.4  \\
 2 & 2  & 1  & 1  & 1/2  & 54  & 1.0  \\
 2 & 2  & 1  & 1  & 3/2  & 34  & 0.7  \\
 2 & 2  & 2  & 0  & 1/2  & 54  & 3.3  \\
 2 & 2  & 2  & 1  & 1/2  & 84  & 10.1  \\
 0 & 2  & 2  & 0  & 1/2  & 42  & 1.4  \\
 0 & 2  & 2  & 1  & 1/2  & 100  &32.9  \\
 0 & 2  & 2  & 1  & 3/2  & 54  &  4.9  \\
 2 & 0  & 2  & 0  & 1/2  & 94  &  21.3  \\
 2 & 0  & 2  & 1  & 1/2  & 74  &  14.0  \\
 2 & 0  & 2  & 1  & 3/2  & 44  &  4.9  \\
\hline
\end{tabular}
\end{center}
\end{scriptsize}
\end{minipage}
\end{table}

\begin{table}
\vspace*{-10.70cm}
\hspace*{7cm}
\begin{minipage}[t]{7cm}
\caption{\label{tab5} As table \ref{tab3} for the $5/2^{-}$ state
of $^{17}$Ne.}
\begin{scriptsize}
\begin{center}
\begin{tabular}{|cccccc|c|}
\hline
 $\ell_x$ & $\ell_y$  & $L$ & $s_x$ & $S$ & $K_{max}$ & $W(^{17}\mbox{Ne})$  \\
\hline
 1 & 1  & 1  & 1  & 3/2  & 62    & 6.8  \\
 1 & 1  & 2  & 1  & 1/2  & 52    & 4.2  \\
 1 & 1  & 2  & 1  & 3/2  & 62    & 7.3  \\
 2 & 2  & 2  & 0  & 1/2  & 84    & 9.1  \\
 0 & 2  & 2  & 0  & 1/2  & 102   & 47.6  \\
 2 & 0  & 2  & 0  & 1/2  & 102   &  25.0  \\
\hline
 1 & 1  & 2  & 0  & 1/2  & 32  & 1.0 \\
 1 & 1  & 2  & 1  & 1/2  & 32  & 2.1 \\
 2 & 2  & 1  & 1  & 3/2  & 64 &  5.8  \\
 2 & 2  & 2  & 0  & 1/2  & 64  & 3.6  \\
 2 & 2  & 2  & 1  & 1/2  & 74 & 10.6  \\
 2 & 0  & 2  & 0  & 1/2  & 42  & 3.1  \\
 2 & 0  & 2  & 1  & 1/2  & 82 & 29.1  \\
 2 & 0  & 2  & 1  & 3/2  & 62  & 7.0  \\
 0 & 2  & 2  & 0  & 1/2  & 82  & 16.6  \\
 0 & 2  & 2  & 1  & 1/2  & 82  & 14.2  \\
 0 & 2  & 2  & 1  & 3/2  & 62  & 6.6  \\
\hline
\end{tabular}
\end{center}
\end{scriptsize}
\end{minipage}
\end{table}

\begin{table}
\begin{minipage}[t]{7cm}
\vspace*{1cm}
\caption{\label{tab6} As table \ref{tab3} for the $7/2^{-}$ state
of $^{17}$Ne. }
\begin{scriptsize}
\begin{center}
\begin{tabular}{|cccccc|c|}
\hline
 $\ell_x$ & $\ell_y$  & $L$ & $s_x$ & $S$ & $K_{max}$ & $W(^{17}\mbox{Ne})$  \\
\hline
 2 & 2  & 4  & 0  & 1/2  & 44 & 9.8  \\
 0 & 4  & 4  & 0  & 1/2  & 54 & 19.4  \\
 4 & 0  & 4  & 0  & 1/2  & 54 & 14.6  \\
 1 & 3  & 3  & 1  & 1/2  & 54 & 20.2  \\
 1 & 3  & 3  & 1  & 3/2  & 44 & 7.0  \\
 3 & 1  & 3  & 1  & 1/2  & 54 & 21.5  \\
 3 & 1  & 3  & 1  & 3/2  & 44 & 7.5  \\
\hline
 2 & 2  & 3  & 0  & 1/2  & 124 & 31.5  \\
 2 & 2  & 3  & 1  & 1/2  & 94 &  10.6  \\
 2 & 2  & 3  & 1  & 3/2  & 94 &  14.9  \\
 2 & 2  & 4  & 0  & 1/2  & 64  & 11.1  \\
 2 & 2  & 4  & 1  & 1/2  & 84 &  31.4  \\
\hline
\end{tabular}
\end{center}
\end{scriptsize}
\end{minipage}
\end{table}

\begin{table}
\hspace*{7cm}
\begin{minipage}[t]{7cm}
\vspace{-7.75cm}
\caption{\label{tab7} As table \ref{tab3} for the $9/2^{-}$ state
of $^{17}$Ne.}
\begin{scriptsize}
\begin{center}
\begin{tabular}{|cccccc|c|}
\hline
 $\ell_x$ & $\ell_y$  & $L$ & $s_x$ & $S$ & $K_{max}$ & $W(^{17}\mbox{Ne})$  \\
\hline
 2 & 2  & 4  & 0  & 1/2  & 44  &  9.6  \\
 0 & 4  & 4  & 0  & 1/2  & 64  &  18.9  \\
 4 & 0  & 4  & 0  & 1/2  & 64  &  14.2  \\
 3 & 1  & 3  & 1  & 3/2  & 84  &  29.3  \\
 1 & 3  & 3  & 1  & 3/2  & 84  &  27.3  \\
\hline
 2 & 2  & 3  & 1  & 3/2  & 104   &  56.0  \\
 2 & 2  & 4  & 0  & 1/2  & 64   &  11.2  \\
 2 & 2  & 4  & 1  & 1/2  & 84   &  30.4  \\
 2 & 2  & 4  & 1  & 3/2  & 64   &  1.9  \\
\hline
\end{tabular}
\end{center}
\end{scriptsize}
\end{minipage}
\end{table}

The lowest valence space for the two protons in $^{17}$Ne is the
$sd$-shell. Therefore the expected dominating components are
$\ell_x$=0,2, $\ell_y$=0,2. By use of these components and 
the core-spin and parity of 1/2$^-$, we can construct three-body
states with total angular momentum and parity $J^{\pi}$=$1/2^-$,
$3/2^-$, $5/2^-$, $7/2^-$, and $9/2^-$.  Nevertheless, it is only
after a full calculation that it is possible to determine precisely
which components are needed to reach a given accuracy. For  the states
mentioned above the components are specified in tables \ref{tab3} to \ref{tab7}.

\begin{figure}
\begin{center}
\vspace*{-1.1cm}
\epsfig{file=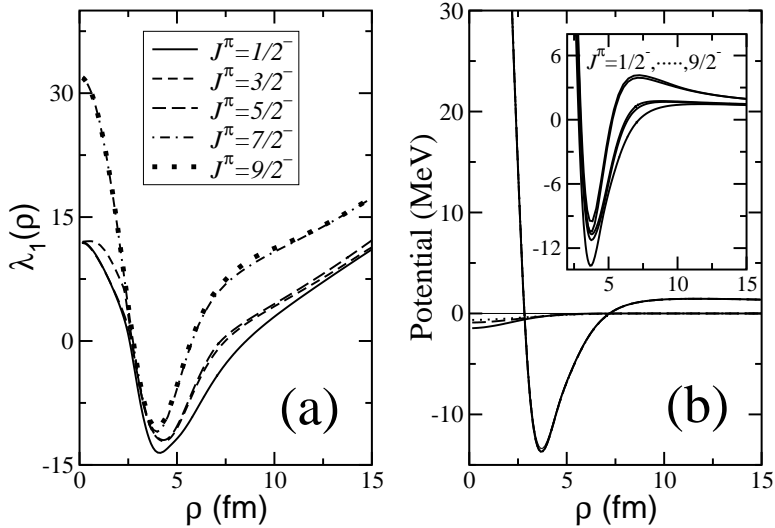,scale=0.40, angle=270}
\end{center}
\caption{\label{fig1} a) The lowest angular eigenvalues 
$\lambda_1(\rho)$ for the $1/2^-$, $3/2^-$, $5/2^-$, $7/2^-$ and
$9/2^-$ states of $^{17}$Ne as function of $\rho$.
b) Outer part: Three different effective three-body 
gaussian potentials that reproduce the experimental two-proton separation 
energy of the $1/2^-$ state in $^{17}$Ne. Their parameters are given 
in table \ref{tab8}. The curve diverging at the origin 
is the corresponding effective potentials in Eq.(\ref{eq4}).  
Inner part: Comparison of the deepest effective 
potential (Eq. (\ref{eq4})) for the different $J^-$ states of $^{17}$Ne. 
The depth of the effective potentials decreases when $J$ increases.}
\end{figure}

\subsection{Effective potentials}

In Fig.\ref{fig1}a we show the deepest angular eigenvalue $\lambda_1$
obtained from equation (\ref{eq2}) for each of the $1/2^-$, $3/2^-$,
$5/2^-$, $7/2^-$, and $9/2^-$ states in $^{17}$Ne.  Their main
characteristics are the hyperspherical values $K(K+4)$ at $\rho$=0, a
minimum due to the attractive short-range interaction and a linear
increase at asymptotically large distances due to the repulsive
Coulomb interaction \cite{fed96}. The $\lambda$-function starting at
0 ($K$=0) is not present due to the use of the $s$-wave phase equivalent 
potential removing the deeply bound $s$-state to account for the Pauli 
principle \cite{gar99}. Details about the higher $\lambda$-functions 
are shown in \cite{gar04}.

From the figure we see that the most attractive $\lambda$ function
in $^{17}$Ne corresponds to the $1/2^-$ state. This agrees with the
fact that the $1/2^-$ state is the only bound state in $^{17}$Ne with
a measured two-proton separation energy of $-944$ keV. Nevertheless,
after the computation, the resulting $^{17}$Ne ground state binding
energy is $-0.79$ MeV, i.e. 150 keV less bound than the experimental
value. This is actually expected, since three-body calculations using
pure two-body interactions typically underbind the system. This
problem is solved by inclusion of the weak effective three-body
potential $V_{3b}$ in Eq.(\ref{eq3}), that accounts for three-body
polarization effects arising when the three particles all are close to
each other.  Therefore the three-body potential has to be of short
range, and its precise shape is not very relevant, since the overall
three-body structure is unchanged. This construction ensures that the
two-body resonances remain unaffected within the three-body system
after this necessary fine-tuning. The effective total potential
entering is then given by Eq.(\ref{eq4}).

In the outer part of Fig.\ref{fig1}b the curves close to the
horizontal zero axis show three gaussian three-body forces with
different ranges and strengths adjusted to reproduce the measured
two-neutron separation energy of 0.94 MeV for the ground state of
$^{17}$Ne.  The corresponding effective total potentials given by
Eq.(\ref{eq4}) are also shown in Fig.\ref{fig1}b, but they can not be
distinguished from each other within the thickness of the curves. This
is because the three-body force is very weak compared to the depth of
the full potential and it is furthermore largest for small
$\rho$-values, where the total potential is highly repulsive.  It is
then clear that the main structure of the system can not be
significantly modified by the choice of one or another of such
three-body interactions. In the inner part of Fig.\ref{fig1}b we show
the computed effective potentials (\ref{eq4}) for the five computed
states of $^{17}$Ne.  The depth of the potential decreases with
increasing total angular momentum. Higher $J$-values correspond then
to higher energies of the $^{17}$Ne state.

\subsection{Spectrum of $^{17}$Ne}

The angular eigenvalues are computed from the effective potentials and
used in the radial equations (\ref{eq3}).  Bound states are obtained
as solutions falling off exponentially at large distances. Resonance
eigenfunctions are found in complete analogy as exponentially falling
solutions to the similar equations obtained after complex rotation of
the hyperradius.

\begin{table}
\caption{\label{tab8}  The 2$^{nd}$ and 3$^{rd}$ columns give the experimental
and computed bound state ($1/2^-$) and decay energies ($3/2^-$,
$5/2^-$, $7/2^-$, $9/2^-$) in $^{17}$Ne (in MeV). The 4$^{th}$ to
6$^{th}$ columns are the strengths of the gaussian three-body forces
that for a range of 3.0 fm, 3.5 fm, and 4.0 fm, give rise to computed
energies in agreement with the experimental values.  The 7$^{th}$
column is the expectation value of the three-body force with range
of 4.0 fm. The last column is the contribution to the norm of the
first three terms in the expansion (\ref{eq1b}) using the three-body
potential of range 4.0 fm.}
\begin{center}
\begin{tabular}{|c|cc|ccc|c|c|}
\hline
$J^{\pi}$ & $E_{exp}$ & $E_{comp}$ & 
      3.0 fm & 3.5 fm  & 4.0 fm  & $\langle V_{3b} \rangle$ (MeV) &
      $\lambda_{n=1,2,3}$ (\%) \\
\hline
 $1/2^-$ &$-0.94$ & $-0.79$ & $-1.5$  & $-0.9$ & $-0.6$ & $-0.2$  &  88.5, 11.1, 0.4 \\
 $3/2^-$ & 0.34   & 0.63    & $-3.4$  & $-2.0$ & $-1.4$ & $-0.3$ &  90.7, 8.9, 0.2 \\
 $5/2^-$ & 0.82   & 0.91    & $-1.1$  & $-0.7$ & $-0.4$ & $-0.1$ &  77.2, 16.9, 5.6 \\
 $7/2^-$ & 2.05   & 2.24    & $-1.8$  & $-1.1$ & $-0.8$ & $-0.2$ &    97.5, 2.3, 0.2 \\
 $9/2^-$ & 2.60   & 2.70    & $-0.1$  & $-0.1$ & $-0.1$ & $-0.1$ &    91.9, 4.2, 3.8 \\
\hline
\end{tabular}
\end{center}
\end{table}

Only the $1/2^-$ state is bound in $^{17}$Ne with a two-proton
separation energy of $-944$ keV. The excitation energies of the four
lowest resonance states are \cite{gui98} 1288$\pm$8 keV,
1764$\pm$12 keV, 2997$\pm$11 keV, and 3548$\pm$20 keV for the $3/2^-$,
$5/2^-$, $7/2^-$, and $9/2^-$ states, respectively.  These excitation
energies correspond to the decay energies (energies above threshold)
shown in the second column of table~\ref{tab8}.  
The computed states (third column) are all slightly underbound
compared to the experiment, but the ordering in the spectrum is
correct.  Additional small attractive three-body potentials are needed
to reproduce the experimental energies. For gaussians we give in
table~\ref{tab8} both the necessary strengths for different ranges and
the expectation values $\langle V_{3b}(\rho) \rangle$ for the
corresponding $J^\pi$ solutions.  The efficiency of the adiabatic
expansion in Eq.(\ref{eq1b}) is seen in the last column of the table.
Already the lowest adiabatic potential accounts for more
than 75$\%$ of the probability and in most cases for more than
90$\%$. The contributions from the different three-body potentials can
hardly be distinguished.

For the ground state of $^{17}$Ne, we give in the last column of table
\ref{tab3} the percentage of the total norm provided by each of the components 
defined by the sequence of quantum numbers in the previous
columns. These numbers are computed with the three-body
potential with range 4 fm, and they do not change significantly when
another choice is made.  When writing the three-body wave function in
the first Jacobi set (upper part of the table) the main contribution
of 89\% is from $s$-waves, and less than 4\% comes from $d$-waves. If
we use any of the other two identical Jacobi sets (lower part of the
table) the $d$-waves contribute with roughly 50\% while 45\% is from
the $s$-waves and the remaining few percents come from $p$-waves. 

It is remarkable that the interference between $s$ and $d$ waves
explicitly included in the calculation plays a negligible
role and falls below the 1\% limit included in the table. The reason
is that rotation of the components from one Jacobi set into another
preserves the value of the total orbital angular momentum $L$.  Then
$sd$ terms ($L$=2) in the second and third Jacobi sets must
correspond to the almost non contributing \{$\ell_x$=$\ell_y=1$, $L$=2,
$S$=3/2\} component in the first Jacobi set, since $L$=2 otherwise
only arises from $\ell_x$=0 or 2 where the antisymmetry dictates zero
spin of the two protons.  Then $L$=2 and the resulting total spin of
$S$=1/2 can not couple to the total angular momentum $1/2$ of the
ground state.  Thus this type of interferences is essentially
excluded.

The unbound excited states of $^{17}$Ne are computed by application of the
complex scaling method. The resonances obtained for $^{17}$Ne are
extremely narrow with widths much smaller than the accuracy of our
calculations.  Actually in \cite{chr02} an experimental lower limit of
26 ps on the lifetime of the two-proton decay of the 3/2$^-$ state is
given.  This lower limit for the lifetime amounts to an upper limit of
$2.5 \cdot 10^{-11}$ MeV for the width of the resonance.  Thus,
application of the complex scaling method allows the use of very small
scaling angles.  Typically complex scaling angles of $\theta=10^{-5}$
are able to find the $^{17}$Ne resonances.  For these scaling angles
the complex scaled $\lambda$'s can hardly be distinguished from the
non-rotated functions in Fig.\ref{fig1}a.  The imaginary parts are very
small and would appear on the zero line if plotted on the figure.

In tables \ref{tab4}$-$\ref{tab7} we show the largest components
included in the calculation of the $3/2^-$, $5/2^-$, $7/2^-$, and
$9/2^-$ states, respectively. For the $3/2^-$ and $5/2^-$ states (as
for the 1/2$^-$) the $s$, $p$, and $d$-waves are enough to get an
adequate description of the system.  Inclusion of $\ell>2$ waves and
use of extrapolated potentials does not significantly modify the
results.  The $s$-wave interaction between proton and core enters only
in combinations with the $d$-wave.  The component $\{ \ell_x$=0,
$\ell_y$=0, $L$=0 \} is simply not allowed in the $5/2^-$ state (the
three intrinsic spins of 1/2 can not couple to 5/2).  For the $3/2^-$
state rotation of the component $\{\ell_x$=$\ell_y$=$L$=0, $s_x$=1,
$S$=$3/2\}$ from the second and third Jacobi sets into the first one
contributes very little through the coupling of $\ell_x$=1 and $\ell_y$=1 due
to antisymmetry and angular momentum conservation.

Contrary to the ground state, the $sd$-interferences contribute
substantially for both the $3/2^-$ and $5/2^-$ wave functions.  For
the $3/2^-$ state they give more than 70\% of the probability in the
first Jacobi set, and about 80\% in the other two, see
table~\ref{tab4}.  For the $5/2^-$ state these probabilities are
roughly 73\% in the first Jacobi set and 77\% in the second and third,
see table \ref{tab5}.  In both states the $p$-waves have a significant
contribution only in the first Jacobi set, while in the other two sets
all the possible \{$\ell_x$= $\ell_y$=1\} components give less than
3\% of the probability.

For $J^\pi$=7/2$^-$ and $9/2^-$ (tables \ref{tab6} and \ref{tab7})
$s$-waves do not contribute if only $s$, $p$, and $d$-waves are
included.  For $J^\pi$=$9/2^-$ the $sd$-interferences are not allowed.
For the $7/2^-$ state in principle the $sd$-interference could
contribute only in the second and third Jacobi sets ($L$=2 and $S$=3/2
can couple to 7/2).  However, rotation into the first Jacobi set
preserves both the $L$=2 and $S$=3/2 values implying $s_x$=1 and
$l_x$=1 due to antisymmetry. These components give a contribution to
the norm smaller than 1\%.  For these two resonance states only
proton-core $d$-waves for both $l_x$ and $l_y$ are significant and
appearing in the lower part of tables \ref{tab6} and \ref{tab7}.

When higher waves are included they do not contribute above the 1\%
level in the second and third Jacobi sets. The reason is simply that 
it is too expensive to increase the
centrifugal barrier. However, the components in the first Jacobi set
follow by rotation. In particular the large components with total
orbital angular momentum $L$=3 and $S$=3/2 must correspond to equally
large contributions of odd partial waves in the first Jacobi set,
which necessarily has $\ell_x\geq 3$.  For $9/2^-$ this is in fact the
dominating component. The same happens with the $L$=4 and $S$=1/2
components, that after rotation into the first Jacobi set give
important contributions corresponding to $sg$-interferences.

\begin{figure}
\begin{center}
\vspace*{-1.1cm}
\epsfig{file=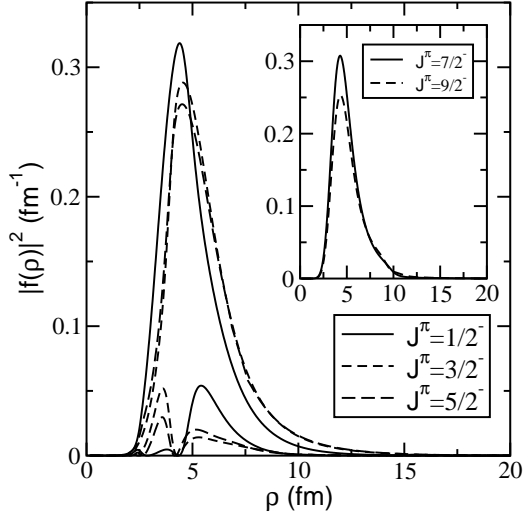,scale=0.4, angle=270}
\end{center}
\caption{\label{fig2} Outer part: Square of the radial wave functions 
corresponding to the two lowest effective potentials $\lambda_n(\rho)$
for the $1/2^-$ (solid curves), $3/2^-$ (short-dashed curves), and
$5/2^-$ (long-dashed curves) states of $^{17}$Ne. Inner part: The
same as the outer part but for the lowest effective potential
$\lambda_n(\rho)$ of the $7/2^-$ (solid curve) and $9/2^-$ (dashed
curve) states.}
\end{figure}

\subsection{Spatial distribution}

In Fig.\ref{fig2} we show the square of the radial wave functions in
the expansion (\ref{eq1b}). This allows a direct comparison of the
contribution of each term to the total wave function. In the outer
part of the figure we show the first two terms in the expansion for
the states with $J^\pi$=1/2$^-$, 3/2$^-$, and $5/2^-$. For
$J^\pi$=7/2$^-$ and 9/2$^-$, shown in the inner part, only the
contribution of the dominating term in Eq.(\ref{eq1b}) is visible.

\begin{figure}
\begin{center}
\epsfig{file=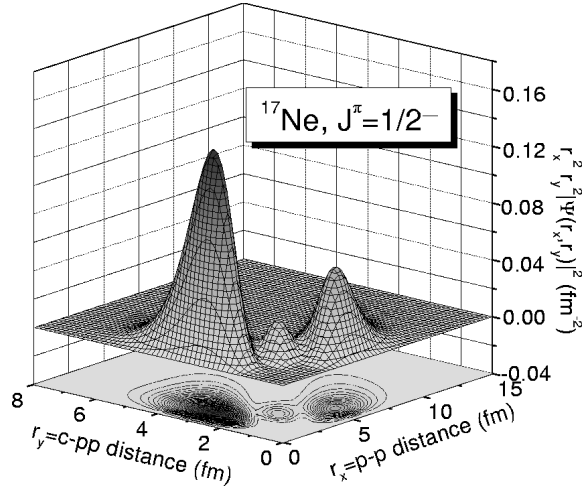,scale=0.35, angle=270}
\end{center}
\caption{\label{fig3} Contour diagram for the probability distribution 
of the $1/2^-$ ground state of $^{17}$Ne. The square of the three-body
wave function is integrated over the directions of the two Jacobi 
coordinates.}
\end{figure}

The complex scaling method provides bound states and resonances
simultaneously as solutions of the complex rotated equation (\ref{eq3})
that asymptotically decrease exponentially. The radial solutions in 
Fig.\ref{fig2} show this asymptotic behaviour already at
distances of around 9 fm.

For the excited states of $^{17}$Ne, computed by the complex scaling
method, the imaginary parts of the radial functions are very small and
only the real parts are visible in the figure. This is a direct
consequence of the extremely small widths of these resonances which
allow a very small complex scaling angle resulting in almost real wave
functions.  As for any resonance, the true radial wave function is
non-square integrable, diverging at large distances as
$\exp{(|\kappa|\rho \sin\beta_R)}$, where $\beta_R$ is the argument of
the resonance. Due to the extremely small values of $\beta_R$ for the
present cases of $^{17}$Ne this divergence has no effect for
$\rho$-values up to several hundred fm.  

Therefore, up to distances around 40-50 fm we can treat the resonances
as bound states, and compute observables like $\langle \rho^2\rangle
$, $\langle x^2 \rangle $ or $\langle y^2\rangle $, that give
information about the spatial distribution of the three-body system.
For each state, we give in table \ref{tab9} such characteristic root
mean square distances between the particles.  The root mean square
radius of the whole three-body system $\langle r^2 \rangle^{1/2}$ in
the $1/2^-$ ground state is 2.8 fm consistent with the experimental
value of 2.75$\pm$0.07 fm given in \cite{oza94}.

\begin{table}
\caption{\label{tab9} For the different computed states of $^{17}$Ne, 
root mean square distances $\langle r^2_{pp} \rangle^{1/2}$, $\langle
r^2_{c,pp} \rangle^{1/2}$, $\langle r^2_{cp} \rangle^{1/2}$, $\langle
r^2_{p,cp} \rangle^{1/2}$, $\langle \rho^2 \rangle^{1/2}$, and
$\langle r^2 \rangle^{1/2}$, where $p$ and $c$ denote a halo proton
and the core, respectively. All the distances are given in fm.  The
results do not depend on the three-body interaction chosen from table
\ref{tab8}.}
\begin{center}
\begin{tabular}{|c|cccccc|}
\hline
 $J^{\pi}$  & $\langle r^2_{pp} \rangle^{1/2}$ & $\langle r^2_{c,pp} 
\rangle^{1/2}$ & 
     $\langle r^2_{cp} \rangle^{1/2}$ & $\langle r^2_{p,cp} \rangle^{1/2}$  &
     $\langle \rho^2 \rangle^{1/2}$   &  $\langle r^2 \rangle^{1/2}$ \\
\hline
 $1/2^-$ & 4.5  & 3.2 & 3.9 & 3.9 & 5.3 & 2.8 \\
 $3/2^-$ & 5.3  & 3.4 & 4.3 & 4.4 & 5.9 & 2.9 \\
 $5/2^-$ & 5.5  & 3.5 & 4.3 & 4.4 & 5.9 & 2.9 \\
 $7/2^-$ & 5.6  & 3.5 & 4.3 & 4.5 & 6.0 & 3.0 \\
 $9/2^-$ & 5.7  & 3.6 & 4.4 & 4.5 & 6.1 & 3.0 \\
\hline
\end{tabular}
\end{center}
\end{table}

The structures derived from these mean values are crudely speaking
triangles where the distance between the two protons is slightly
larger than the distance between proton and core. The dimension increases 
with angular
momentum. A more detailed view is obtained in the contour diagram in
Fig.\ref{fig3}. We plot the square of the three-body wave function
multiplied by the phase space factors and integrated over the
directions of the two Jacobi coordinates. We observe that a prominent
peak is found for distances between the two protons of about 2 fm and
the two-proton center of mass distance to the core of about 3
fm. Another smaller peak is found for corresponding distances of about
5 fm and 1 fm, respectively. These two shapes can be described as two
protons either on the same side of the core (at some distance apart)
or at almost opposite sides of the core. The structures of the excited
states are very similar apart from the dimension increasing with $J$.

The distribution overlaps to some extent the core density with the
potential danger of violating the basic few-body assumptions. 
However, an assessment of model validity is 
only possible by comparing many predictions with measured observables.

The size of a three-body system is for quantum halos related to the
binding energy. The repulsive Coulomb potential may provide a very
small binding energy but at the same time confining the structure
spatially. The characteristic halo features are revealed by the
hyperradial extension measured in units of an appropriate length scale
$\rho_0$, designed in \cite{jen03} to measure the content
of non-classical probability. In our calculation we have 
$\rho_0 \approx 5$ fm, which implies that the
dimensionless measure of the binding energy $m B \rho^2_0 /\hbar^2$ is
around 0.6 and the mean square hyperradius in units of $\rho^2_0$ is
about 1.2. This wave function is therefore to a large extent in the
classically allowed region and the properties cannot be obtained by
scaling relations independent of the details of the interactions. The
$^{17}$Ne system is not a quantum halo as defined in \cite{jen03}
although the structure still can be described as a three-body system.

\subsection{Widths of the resonances}

The positive energies of the excited states imply that they can decay
into three separate particles and also to lower-lying states
by electromagnetic transitions. However, the widths of the resonances
are very small and beyond the accuracy of the otherwise very efficient
complex scaling method. Fortunately the hyperspherical adiabatic
method provides generalized radial potentials which are responsible
for the three-body bound state structure as well as the asymptotic
behavior of the continuum wave functions. The decay widths into
separate free particles can then be estimated in the WKB
approximation.  This is especially tempting since about 90\% of the
wave function is described by the lowest adiabatic potential and
since the widths are small and the effective barrier is smooth.
In \cite{fed96} the accuracy of the WKB approximation was estimated
for a system of three charged particles. The WKB turned out to be 
accurate to within a factor of five including the uncertainty from 
the preexponential factor, the ``knocking rate".

\begin{figure}
\begin{center}
\vspace*{-1.1cm}
\epsfig{file=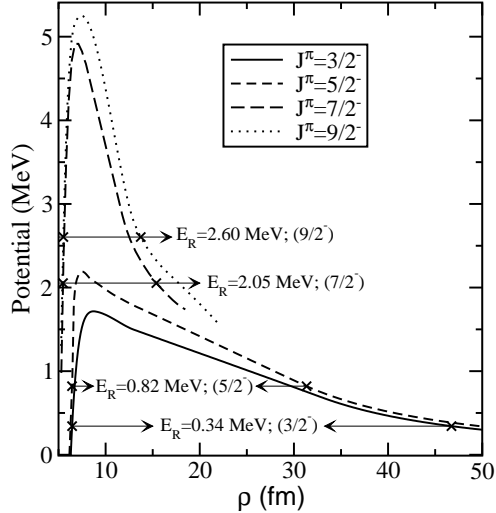,scale=0.4, angle=270}
\end{center}
\caption{\label{fig4} Barrier of the most contributing effective 
potential for the $J^\pi$=$3/2^-$, $5/2^-$, $7/2^-$, and $9/2^-$
resonances in $^{17}$Ne. The crosses show the turning points and the
arrows show the energies corresponding to each of the resonances. }
\end{figure}

This use of the WKB approximation, where the lowest adiabatic
potential provides the tunneling barrier, is a rather subtle
application. There is a striking resemblance with computation of
$\alpha$-decay widths from tunneling through centrifugal and Coulomb
barriers. However, $\alpha$-decay is assumed to be a two-body process
where the $\alpha$-particle is formed with some probability and then
tunnels through the two-body barrier.  The two-proton decay is a
three-body problem and it is only due to the efficiency of the
hyperspheric expansion that one suitable generalized potential becomes
available. The tunneling probability through the corresponding barrier
is, in the one $\lambda$-approximation, then precisely describing the
resonance width for decay into the corresponding channel of three
particles in the final state. On this level there is a complete
equivalence with the $\alpha$-decay computation. The WKB expression is
then a rather good approximation to the full three-body coupled
channel problem under the two conditions that the lowest adiabatic
hyperspheric potential and the WKB-approximation both provide
sufficient accuracy.

We therefore have to compute the transmission coefficient through the
barrier of the lowest effective potential $V_{\mbox{eff}}$ (see
Eq.(\ref{eq4})) constructed by using the dominating angular
eigenvalue, see Fig.\ref{fig1}b. If crossings of potentials occur
along the way through the barrier we follow the path determined by
minimum change of the wave function. This corresponds to a path along
the most smooth adiabatic potential in full agreement with the basic
assumptions of the WKB approximation.  Thus the WKB transmission
coefficient is estimated by
\begin{equation}
T=\exp{\left\{-2 \int_{\rho_i}^{\rho_o} \left[    
\frac{2 m}{\hbar^2}(V_{\mbox{eff}}(\rho)-E_R)
                           \right]^{1/2} d\rho \right\} } \; ,
\end{equation}
where $E_R$ is the energy of the resonance, and $\rho_i$ and $\rho_o$
are the inner and outer classical turning points defining the distance
through the barrier. In Fig.\ref{fig4} we show the barriers of the
resulting effective potential for the four resonances in $^{17}$Ne
with $J^{\pi} = 3/2^-, 5/2^-, 7/2^-$ and $9/2^-$. The outer turning
points of the effective potentials are located at distances of around 15
fm for the $7/2^-$ and $9/2^-$ states, above 30 fm for the $5/2^-$
resonance, and above 45 fm for the $3/2^-$ resonance. These
potentials, and therefore the angular eigenvalues, must then be
computed accurately below these values. Inaccuracies typically produce
too high eigenvalues and therefore too large barriers with too small
transmission coefficients or equivalently too small widths.

\begin{table}
\caption{\label{tab10} Turning points, frequency, transmission coefficients, 
average lifetimes, and resonance widths for the $3/2^-$, $5/2^-$, $7/2^-$, 
and $9/2^-$ resonances in $^{17}$Ne. $\rho_i$ and $\rho_o$ are given in fm,
$f$ in MeV/$\hbar$, $t_a$ in s, and $\Gamma$ in MeV. }
\begin{center}
\begin{tabular}{|ccccccc|}
\hline
$J^{\pi}$ & $\rho_i$  & $\rho_o$ & $f$ & 
                                T  & $t_{a}$ & $\Gamma$ \\
\hline
 $3/2^-$ & 6.4  & 46.7 & 3.90 &$9.2\cdot10^{-13}$ & $1.8\cdot10^{-10}$  
                                            & $3.6\cdot10^{-12}$ \\
 $5/2^-$ & 6.4  & 31.3 & 3.82 &$3.3\cdot10^{-11}$ & $3.6\cdot10^{-12}$  
                                            & $1.3\cdot10^{-10}$ \\
 $7/2^-$ & 5.5  & 15.4 & 1.54 &$3.7\cdot10^{-3}$  & $1.2\cdot10^{-19}$  
                                            & $5.6\cdot10^{-3}$ \\
 $9/2^-$ & 5.5  & 13.7 & 1.59 &$3.3\cdot10^{-3}$  & $1.2\cdot10^{-19}$  
                                            & $5.3\cdot10^{-3}$ \\
\hline
\end{tabular}
\end{center}
\end{table}

Once the transmission coefficient is computed, the decay constant can
be obtained as $\Gamma /\hbar$=$f T$, where the frequency $f$ is given
in terms of the radial extension $\rho_i$ of the attractive pocket and
the velocity $v$ of the particle moving in it, i.e. $f$=$v/\rho_i$.  We
estimate the velocity $v$ by equating the kinetic energy ($m v^2/2$)
and the maximum potential energy $E_R$+$|V_{\mbox{eff}}^{(min)}|$, where
$|V_{\mbox{eff}}^{(min)}|$ is the maximum depth of the effective
potential. This estimate of the knocking rate can easily be
off by a factor of two. Furthermore this computation implicitly
assumes that the preformation factor is unity, i.e. that the decaying
state really is a three-body state consisting of two protons outside
the $^{15}$O core.

Finally, the average lifetime $t_a$ or the width $\Gamma$ of the
resonances are found by inverting the decay constant or multiplying by
$\hbar$, respectively.  In table~\ref{tab10} we summarize the results
for the different resonances.  The computed width of the $3/2^-$
resonance is consistent with the upper limit of $2.5
\cdot 10^{-11}$ MeV quoted in \cite{chr02}. This limit is in fact 
essentially theoretical as obtained in a shell model calculation for the
magnetic dipole deexcitation probability. The lack of protons in the
experiment then leads to the conclusion that the width must exceed the
computed electromagnetic value.

Similarly a lower limit for the two-proton decay of the $5/2^-$
resonance is deduced to be about $3 \cdot 10^{-10}$ MeV \cite{chr02}
which is about three times larger than our crude calculation shown in
table~\ref{tab10}.  The two higher-lying resonances have much larger
widths of about 5~keV due to the higher energies and the rather narrow
barriers. This behavior can not be predicted from angular momentum and
energy systematics. The three-body model is indispensable, not only
because the final state asymptotically consists of three particles but
more interestingly because the short and intermediate distances
determine the widths.

\section{Accuracy of model results}

The calculated results can not be more reliable than the model.
The basic assumptions are that the active few-body degrees of
freedom are responsible for all the computed properties. The intrinsic
constituent particle degrees of freedom are only implicitly treated
through the properties of the two-body interactions. Polarization
effects are included but not explicit contributions from intrinsic
structure. Therefore cluster models are most suited in descriptions of
halo systems while individual applications beyond halos still can be
rather accurate if the different degrees of freedom are sufficiently
weakly coupled.

Ultimately only practical tests can decide on the issue of model
reliability. However, a necessary prerequisite is a series of accuracy
tests of the model itself. This means that the numerical methods must
reach convergence to the requested level of accuracy. This is
not trivial in the present type of cluster computations. Another
source of uncertainty is the parametrization of the interactions,
e.g. spin-dependence, radial form factors and the numerical values of 
the parameters. We shall in this section report on these tests.

\subsection{Convergence and basis expansion}

The convergence of the calculations must be reached at three different
levels. First, in the hyperspheric adiabatic expansion of
Eq.(\ref{eq1b}) the number of included potentials must be sufficient.
This was shown in the last column of table~\ref{tab8}. Typically 
three terms are enough to provide contributions to the wave functions 
larger than 99\% of the probability. Sometimes one more adiabatic potential has 
to be included for additional confidence.

Second, the expansion of the angular eigenfunctions $\phi_n^{(i)}$ in 
Eq.(\ref{eq2}) in terms of hyperspherical harmonics has to contain
all the components giving a relevant contribution to the wave
functions. These are the components shown in tables \ref{tab3} to
\ref{tab7}. Together with them, additional components giving contributions
smaller than 1\% have also been included in the calculations.

Third, once the components have been chosen, convergence must be secured 
in the expansion in hyperspherical harmonics by requiring inclusion of a 
sufficiently large number of them, i.e.  ${\cal Y}_{\ell_x
\ell_y, L}^{K}(\alpha_i, \Omega_{x_i}, \Omega_{y_i})$. For each component
we include all these basis functions with hyperspherical quantum
number less than a maximum number $K_{max}$ which may depend on the
quantum numbers of the components.  We should emphasize that the basis
sizes we discuss are for each Faddeev component, i.e. related to each
Jacobi coordinate system. If only one Jacobi set is used a much larger
$K_{max}$ is necessary for convergence.

\begin{figure}
\begin{center}
\vspace*{-1.1cm}
\epsfig{file=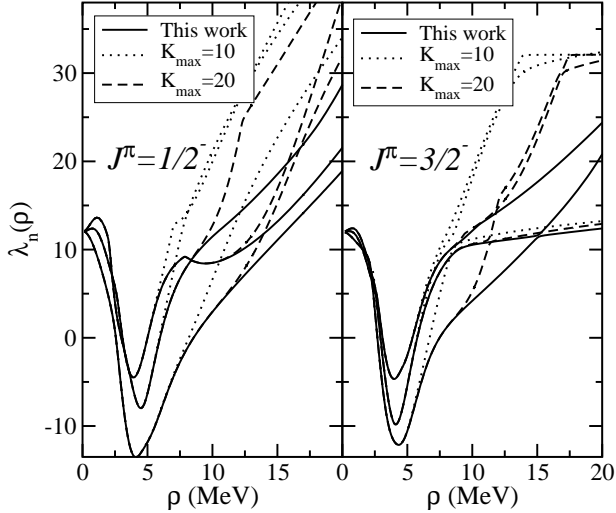,scale=0.4, angle=270}
\end{center}
\caption{\label{fig5} Comparison of the computed $\lambda$ functions 
for different values of $K_{max}$ for the $^{17}$Ne states
$J^\pi=1/2^-$ (left) and $J^\pi=3/2^-$ (right). The solid curves are
the results with $K_{max}$ from tables \ref{tab3} and \ref{tab4}. }
\end{figure}

In the calculations this hypermoment $K_{max}$, given in the sixth
column of tables \ref{tab3}-\ref{tab7}, is by far large enough to
reach convergence in the region of contributing $\rho$-values.  This
is illustrated in Fig.\ref{fig5} where we show the three first angular
eigenvalues, the $\lambda$-functions, for the $1/2^-$ and $3/2^-$
states in $^{17}$Ne for three different values of $K_{max}$.  For the
ground state we observe that at a distance of around 7 fm the lowest
$\lambda$ with $K_{max}$=10 begin to differ from the converged result.
From Fig.\ref{fig2} we see that this significantly can modify the tail
of the $1/2^-$ radial wave function.  In fact, using the three-body
interactions given in table \ref{tab8} the two-neutron separation
energy of the $1/2^-$ state with $K_{max}$=10 is only $-0.58$
MeV, almost 400 keV less bound than in the full calculation.
For $K_{max}$=20 the discrepancy from the converged lowest
$\lambda$ for $J^\pi$=1/2$^-$ begins at distances larger than 10 fm,
and the effect on the radial wave function is not important.  When
$K_{max}$=20 the two-neutron separation energy is $-0.93$ MeV, pretty
close to the one we obtained.  Therefore for the ground state at least
a value of $K_{max}$=20 is needed.

For the first excited state, $3/2^-$, the discrepancies between the
different calculations begin at lower distances, see Fig.\ref{fig5}.
When $K_{max}$=10 the lowest $\lambda$ begins to differ from the other
calculation already at $\rho$=6 fm.  At this value of $\rho$ the
radial wave function in Fig.\ref{fig2} is close to its maximum, and
therefore the results obtained with $K_{max}$=10 are not reliable.
Actually, the $3/2^-$ resonance energy obtained in this case is above
700 keV, more than a factor of two higher than in the full
calculation.  For $K_{max}$=20 the computed $3/2^-$ resonance energy
is still 15\% higher than the 340 keV obtained in the full
calculation.  The same kind of convergence is seen for higher values
of $J$.

We can then conclude that for the excited states of $^{17}$Ne values
of $K_{max}$ higher than 20, typically around 30, are desirable. As
shown in tables \ref{tab4}-\ref{tab7} we used in our
calculations $K_{max}$ values significantly larger than 30 for some of
the components.  This is because the estimates of the widths of the
resonances require accurate effective potentials for much larger
distances than 10~fm, e.g. up to about 50 fm for the $3/2^-$
resonance. This accuracy is achieved with the $K_{max}$ values given
in the tables.  Thus reliable width calculations in a three-body model
require a rather large individual basis for each of the Faddeev
components related to the different Jacobi sets.

\subsection{The shape of the radial potential}
\label{shape}

In this section we investigate the dependence of the results on the
radial shape of the nuclear two-body interactions. The results
discussed so far have been obtained with the proton-proton interaction
(G) from \cite{gar04,gar97} and the Woods-Saxon (WS) proton-core
interaction in Eq.(\ref{eq1}).

Let us consider now also the gaussian (G) proton-core interaction
given in Eq.(\ref{eq5}) and table~\ref{tab1} and the sophisticated
Argonne (A) proton-proton interaction $v_{18}$
described in \cite{wir95}. This is a nonrelativistic potential that
reproduces $pp$ and $np$ scattering data for energies from 0 to 350
MeV, $nn$ low-energy scattering data, and the deuteron properties.
Different combinations of these potentials permit comparison and
reliability tests of the results.  The calculations are denoted 
G+WS, A+WS, G+G, ..., where the first and second labels refer to the
proton-proton and proton-core interactions, respectively.

\begin{figure}
\begin{center}
\vspace*{-1.1cm}
\epsfig{file=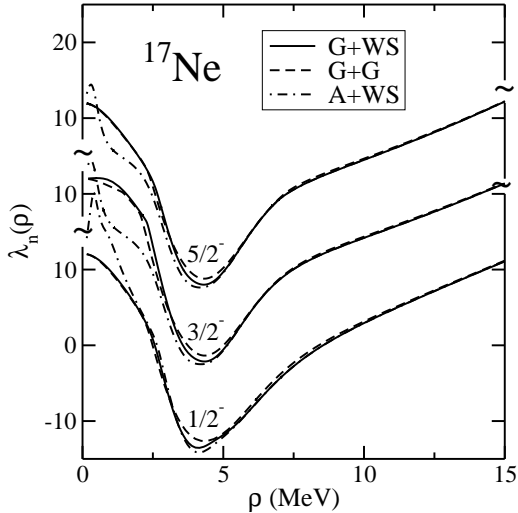,scale=0.40, angle=270}
\end{center}
\caption{\label{fig6} Comparison between the deepest $\lambda$-functions
for the $1/2^-$, $3/2^-$, and $5/2^-$ states in $^{17}$Ne when 
different proton-proton and proton-core interactions are used (G=Gaussian, 
WS=Woods-Saxon, A=Argonne). In the label the first symbol
refers to the proton-proton potential and the second to the 
proton-core. }
\end{figure}

We have repeated all the three-body calculations for the G+G and
A+WS cases.  The decisive $\lambda$-functions entering in the radial
equations (\ref{eq3}) are compared in Fig.\ref{fig6} for the $1/2^-$,
$3/2^-$, and $5/2^-$ states in $^{17}$Ne.  At large
distances the $\lambda$-functions are basically identical in the three
calculations, since the low energy properties of the two-body
potentials are the same. At short distances some differences are visible.  
The G+G calculations produce a slightly less deep pocket in the potentials 
than in the other two cases.  This means that the
phenomenological three-body potentials needed to fit the measured
energies of the $^{17}$Ne states are also slightly different.

\begin{figure}
\begin{center}
\vspace*{-1.1cm}
\epsfig{file=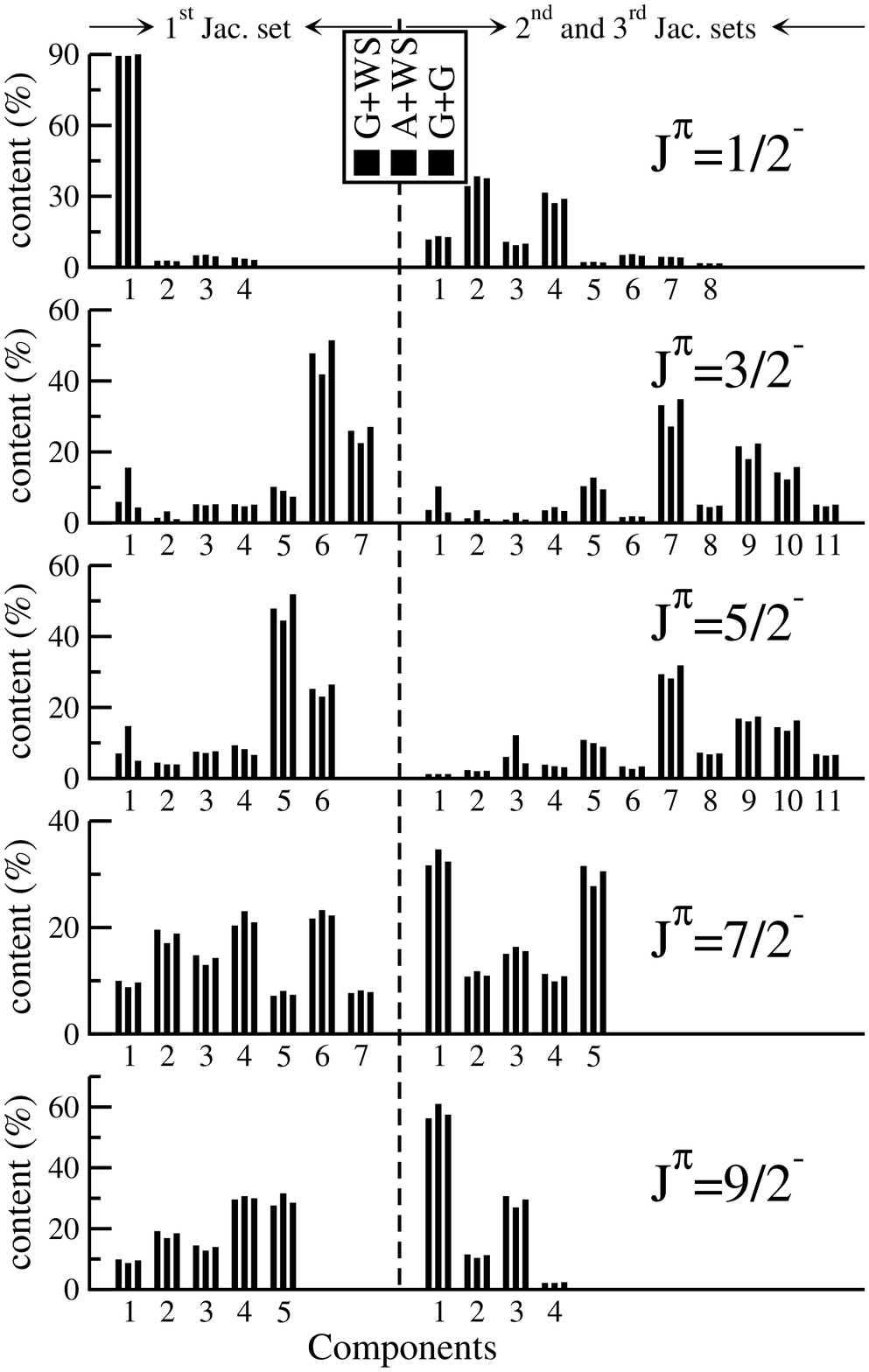,scale=0.45}
\end{center}
\caption{\label{fig7} Comparison of the weights of the 
components for the $^{17}$Ne states. The left part
refers to the first Jacobi set and the right to the second and
third Jacobi sets. The components are ordered as in 
tables~\ref{tab3}-\ref{tab7}. For each component the three
bars, from left to right, refer to the G+WS, A+WS, and G+G
computations, respectively.}
\end{figure}

In total, after fitting the resonance energies in $^{17}$Ne by use of
gaussian three-body  forces, the results are essentially the same as the
G+WS results previously discussed.  The radial wave functions
are hardly distinguishable from the ones shown in Fig.\ref{fig2}. 
In Fig.~\ref{fig7} we compare the weights of the components for the 
three calculations and the five computed states in $^{17}$Ne. The 
components are ordered as in tables~\ref{tab3} to \ref{tab7}. The
left part of the figure corresponds to the components when the 
full three-body wave function is written in the first
Jacobi set, and the right part to the ones in the second and third
Jacobi sets. The three computations give very similar results. For
each state the dominant components are the same in all the three 
cases, and the overall three-body structure is essentially unchanged.

We can then conclude that details like the radial shapes of the
two-body interactions are not changing the properties of the
three-body system within the accuracy provided by the inherent
accuracy of the model.

\section{Previous works}

Different theoretical works concerning the structure of $^{17}$Ne 
can be found in the literature. Most of them
\cite{for01,nak98,mil97} focus only on the ground state structure and in
particular on the weight distribution of the $s$ and $d$-waves. Only
\cite{tim96} and \cite{gri03} report on  more detailed calculations of 
the $^{17}$Ne spectrum.

In \cite{tim96} the three-cluster generator coordinate model is used
to obtain $^{17}$Ne.  The
proton-proton interaction is adjusted to reproduce the experimental
binding energy.  Although no numbers are quoted, the
$s$-wave contribution is expected to dominate.  The same conclusion is
reached in \cite{nak98} where they analyze the experimental parity
dependence of the Coulomb shifts of the $A$=17 isoquartets.  They used
harmonic oscillator wave functions with a state dependent radius and
the Coulomb potential of a uniformly charged sphere.  The two protons
of the ground state of $^{17}$Ne then seem to occupy the
$s_{1/2}$-orbits.

In \cite{mil97} the $d^2$-configuration is for the first time found to
dominate in the ground state of $^{17}$Ne.  This is concluded by
investigating the $\beta^+$ decay to the first excited state of
$^{17}$F. The decay is found to be roughly a factor of two larger than
expected from the nuclear matrix elements that reproduce the $\beta^-$
decay into $^{17}$N.  
In \cite{for01} the domination of the $d^2$-configuration is confirmed
by investigating the difference in Coulomb energy for $s$ and
$d$-orbitals.  They use Woods-Saxon potentials and an uniformly
charged sphere.  A coupling of either two $s_{1/2}$ or two
$d_{5/2}$-protons to the $A$=15 core immediately reveals that the
$d^2$-component dominates.  To reproduce the measured $^{17}$Ne
two-proton separation energy around 78\% of the wave function must be
of $d^2$-character.

\begin{table}
\caption{\label{tab11} Computed $s$, $p$, and $d$ wave contents in the
$^{17}$N (upper part) and $^{17}$Ne (lower part) wave
functions. The
results in parenthesis are the ones obtained in ref.\cite{gri03}.}
\begin{center}
\begin{tabular}{|c|ccc|}
 \hline
 $J^{\pi}$  &  $s$  &  $p$  &  $d$   \\
 \hline
$1/2^-$ & 38.9 (39.8) & 6.7 (4.5) & 54.4 (55.6) \\
$3/2^-$ & 32.8 (36.8) & 2.2 (3.0) & 65.0 (60.0) \\
$5/2^-$ & 32.9 (34.5) & 4.2 (3.2) & 62.9 (61.6) \\
\hline
$1/2^-$ & 45.4 (48.1) & 5.6 (4.0) & 49.0 (47.8) \\
$3/2^-$ & 39.2 (38.1) & 2.1 (2.8) & 58.7 (58.4) \\
$5/2^-$ & 37.4 (36.2) & 3.3 (3.0) & 59.3 (60.1) \\
 \hline
\end{tabular}
\end{center}
\end{table}

A detailed calculation of $^{17}$Ne and $^{17}$N has been reported
recently in \cite{gri03}.  A three-body model is employed and
Woods-Saxon potentials between the three pairs of particles are used.
The weight of the different partial waves for the ground state
contains around 40\% and 48\% $s$-wave in the $^{17}$N and $^{17}$Ne
wave functions, respectively.  These values decrease for the first and
second excited states, where the $s^2$-wave content varies from case
to case between 34.5\% and 38.1\%.  In table \ref{tab11} we compare
our results to the ones given in \cite{gri03}. For completeness we also
include our results for the $^{17}$N case. The agreement is
rather good in all the cases.  However, apparently in \cite{gri03}
interferences between $s$ and $d$ waves are not considered, while we
have found that for the excited states they are crucial.  Furthermore,
it is not easy to understand how two protons in the $s_{1/2}$-waves
can give rise to the excited $5/2^-$ state, provided the $^{15}$O core
still has spin of 1/2. It is also important to remember that by
$\ell$-wave content we refer to the weight of the components in tables
\ref{tab3}$-$\ref{tab7} containing that value of $\ell$. In this way, 
our $\ell_x$=$\ell_y$=0 components are not fully equivalent to the
$s^2$-content in the shell model sense.

The spectrum of $^{17}$Ne is previously investigated only
in \cite{tim96} and \cite{gri03}.  In \cite{tim96} the nucleon-nucleon
interaction is adjusted to reproduce the $^{17}$Ne ground state
binding energy, but the same interaction is not able to describe the
$^{16}$N and $^{16}$F spectra.  Even the level order is wrong and 
the 3$^-$ level comes out as the ground state in both cases.
For the three-body systems the ground state angular momentum is
correct in \cite{tim96}, but the 3/2$^-$ and 5/2$^-$ levels are
reversed compared to the experimental data.  The lack of consistency
between the two-body interaction needed to reproduce the three-body
ground state and the one reproducing the two-body spectra
of the two-body subsystems is certainly a weak point in this
calculation.

The same deficiency is present in \cite{gri03}, where the three-body
structure in $^{17}$Ne obtained with the two-body interactions that
reproduce the $^{16}$F spectrum also reverse the positions of the
$3/2^-$ and $5/2^-$-states.  The $5/2^-$-level is even bound by
$-0.41$ MeV, while the $3/2^-$-state has an energy above threshold of
0.89 MeV.  A three-body force adjusted for each of the states is
needed to restore the level order and the positions as found in
experiments.  It is then clear that the three-body force is forced to
play a too important role, since the $5/2^-$ energy is moving up from
$-0.41$ MeV to the experimental 0.82 MeV and simultaneously the
$3/2^-$ is moved down in energy.

In the present work these problems are not encountered. When only the
two-body forces describing the proton-proton and $^{16}$F properties
are used, the energies of the $^{17}$Ne states are the ones in the
third column of table~\ref{tab8}. They follow the experimental
ordering, and use of a small effective three-body force is enough to
fit the experimental energies.

The fact that in \cite{gri03} the same inversion as in \cite{tim96}
appears can be related to the inappropriate spin dependent operators
used in the two-body nucleon-core interaction. 
Contrary to the present work the interaction in
\cite{gri03} is mixing the $d_{3/2}$ and the $d_{5/2}$-states used as 
an essential part of the valence space for the outer nucleons. The
low-lying resonances in $^{16}$N and $^{16}$F do not have the
desirable pure $d_{5/2}$-character \cite{gar03}.

Finally the partial two-proton decay widths of the three-body
resonances are estimated in the present work, see table \ref{tab10}.
In \cite{gri03} the widths of the $3/2^-$ and $5/2^-$-states
are found to be $4.1\cdot 10^{-19}$ MeV and $1.2\cdot 10^{-11}$ MeV,
respectively smaller by seven and one order of magnitude compared to
our results. This computed $3/2^-$ value is in apparent agreement with
the experimental limit. However, their computed width of the $5/2^-$
resonance is already at least one order of magnitude too small. The
eight orders of magnitude difference between these two computed values
seem to be in clear disagreement with measurements which indicates
that the lowest and highest of these resonances have widths slightly
below and above the electromagnetic decay widths.

The origin of these discrepancies can not be decided without full
repetition of all computations. Two effects are perhaps important in
this connection. The first is that the basis size has to be rather
large to provide a sufficiently accurate description all the way up to
the relatively large distances where important contributions to the
widths are determined. Too small a basis would usually lead to too
high potential energies and therefore too large barriers and too small
widths. The maximum hypermoment must be very large when only
hyperharmonics in one Jacobi set are used.

The second effect is that the hypersherical adiabatic expansion in
principle includes all decay channels, i.e. also sequential decay
which often is the dominating mode. For the $3/2^-$-state this is
apparently not allowed due to energy conservation. However, a kind of
virtual decay, where the tail of the two-body resonance is exploited,
would enhance the decay probability. This effect is included in
our formulation since the two-body resonance configuration influences
the lower adiabatic potentials to fairly large hyperradii. This effect
is difficult to catch with the direct hyperharmonic expansion
method.

\section{Summary and conclusions}

The structure of the lightest Borromean nucleus $^{17}$Ne is
investigated in details in a three-body model where two protons
surround the $^{15}$O core.  We use the hyperspheric adiabatic
expansion method with two-body interactions adjusted to
reproduce the properties of the two-body subsystems. The
spin-dependent form of the proton-core interaction must be consistent
with the mean-field approximation.  Otherwise the assumed decoupling
of core and valence motion is violated due to the Pauli principle and
the angular momentum conservation.  We use two different radial shapes
for the proton-core potential both consistent with the measured size
of $^{15}$O. The Coulomb potential is derived from a gaussian charge
distribution with the measured root mean square radius.

The three-body ground state and four measured excited states of
$^{17}$Ne are computed.  The structure for the ground state is
found to be about 45\% of $s^2$ ($\ell_x$=$\ell_y$=0) and 50\% of
$d^2$ ($\ell_x$=$\ell_y$=2) proton-core components. For the two lowest
excited states the $sd$ relative proton-core states are dominating
while only roughly 18\% is of $d^2$-character. For the two
highest resonances, $d^2$-components dominate completely in the 
second and third Jacobi sets, while contributions from higher partial 
waves ($sg$ and $pf$-components) can be important in the first Jacobi
set. Properties of these states are not computed in previous works.

The spatial distribution is slightly more extended than for ordinary
nuclei and with a clear tendency to show two components, i.e. one
where the protons and the core are distributed in a triangle and one
where the protons try to be on opposite sides of the core.  The
distribution is not in the classically forbidden region and the
scaling relations specific for quantum halos are not obeyed. The
criteria for quantum halos are not fulfilled. Still the three-body
model seems to give a rather good description of the structure of
these five lowest-lying states. The two-proton decay widths are
computed by use of the WKB approximation applied on the adiabatic
potentials. The computed results are within the experimentally
determined limits.  The width estimates employ a recently established
but non-trivial formulation which relies on the efficiency of both the
hyperspheric adiabatic method and the separation at small and
intermediate distances of the intrinsic and the necessary asymptotic
three-body degrees of freedom.

Accuracy of these computations can be divided into three parts, i.e.  (i)
model assumptions related to its applicability, (ii) input determination,
uniqueness and reliability, (iii) numerical accuracy including convergence
of expansions. First the consistency of the results supports the credibility 
of the model. Second we carefully evaluated the accuracy both by testing for
convergence and by using different two-body interactions.

In conclusion, the three-body model describes efficiently the cluster
structure of $^{17}$Ne. The hyperspheric adiabatic expansion in
the same framework provides consistently also the simple semiclassical
estimates of resonance decays into three-body final states.

\section{acknowledgments}
We want to thank Hans Fynbo for useful suggestions and discussions.

\end{document}